\documentclass[graybox]{svmult}

\usepackage{mathptmx}       
\usepackage{helvet}         
\usepackage{courier}        
\usepackage{type1cm}        
\usepackage{makeidx}         
\usepackage{graphicx}        
\usepackage{multicol}        
\usepackage[bottom]{footmisc}

\usepackage{multirow}
\usepackage{hhline}

\usepackage{url}

\makeindex             

\begin{document}

\title*{Beyond the EULA: Improving consent for data mining}

\author{Luke Hutton and Tristan Henderson}
\institute{Luke Hutton \at Centre for Research in Computing, The Open University, Milton Keynes, MK7 6AA, UK \email{luke.hutton@open.ac.uk}
\and Tristan Henderson \at School of Computer Science, University of St Andrews, St Andrews KY16 9SX, UK \email{tnhh@st-andrews.ac.uk}}
\maketitle

\abstract{
Companies and academic researchers may collect, process, and
distribute large quantities of personal data without the explicit
knowledge or consent of the individuals to whom the data pertains.
Existing forms of consent often fail to be appropriately readable and
ethical oversight of data mining may not be sufficient. This raises
the question of whether existing consent instruments are sufficient,
logistically feasible, or even necessary, for data mining. In this
chapter, we review the data collection and mining landscape, including
commercial and academic activities, and the relevant data protection
concerns, to determine the types of consent instruments used. Using
three case studies, we use the new paradigm of human-data interaction
to examine whether these existing approaches are appropriate.  We then
introduce an approach to consent that has been empirically
demonstrated to improve on the state of the art and deliver meaningful
consent. Finally, we propose some best practices for data collectors
to ensure their data mining activities do not violate the expectations
of the people to whom the data relate.}

\section{Introduction}
\label{s:intro}

The ability of companies to collect, process, and distribute large
quantities of personal data, and to further analyse, mine and generate
new data based on inferences from these data, is often done without
the explicit knowledge or consent of the individuals to whom the data
pertains. Consent instruments such as privacy notices or End User
License Agreements (EULAs) are widely deployed, often presenting
individuals with thousands of words of legal jargon that they may not
read nor comprehend, before soliciting agreement in order to make use
of a service. Indeed, even if an individual does have a reasonable
understanding of the terms to which they have agreed, such terms are
often carefully designed to extend as much flexibility to the data
collector as possible to obtain even more data, distribute them to
more stakeholders, and make inferences by linking data from multiple
sources, despite no obvious agreement to these new practices.

The lack of transparency behind data collection and mining practices
threatens the agency and privacy of data subjects, with no practical way
to control these invisible data flows, nor correct misinformation or
inaccurate and inappropriate inferences derived from linked data.
Existing data protection regimes are often insufficient as they
are predicated on the assumption that an individual is able to detect
when a data protection violation has occurred in order to demand
recourse, which is rarely the case when data are opaquely mined at
scale.

These challenges are not unique to commercial activities, however.
Academic researchers often make use of datasets containing
personal information, such as those collected from social network
sites or devices such as mobile phones or fitness trackers. Most
researchers are bound by an obligation to seek ethical approval from
an institutional review board (IRB) before conducting their research.
The ethical protocols used, however, are inherited from post-war
concerns regarding biomedical experiments, and may not be appropriate
for Internet-mediated research, where millions of data points can be
collected without any personal interventions. This raises the
question of whether existing consent instruments are sufficient,
logistically feasible, or even necessary, for research of this nature. 
 
In this chapter we first review the data collection and mining
landscape, including commercial and academic activities, and the
relevant data protection laws, to determine the types of consent
instruments used.  Employing the newly-proposed paradigm of Human-Data
Interaction, we examine three case studies to determine whether these
mechanisms are sufficient to uphold the expectations of individuals,
to provide them with sufficient agency, legibility and negotiability,
and whether privacy norms are violated by secondary uses of data which
are not explicitly sanctioned by individuals. We then discuss various
new dynamic and contextual approaches to consent, which have been
empirically demonstrated to improve on the state of the art and
deliver meaningful consent.  Finally, we propose some best practices
that data collectors can adopt to ensure their data mining activities
do not violate the expectations of the people to whom the data relate.

\section{Background}
\label{s:bg}

Data mining is the statistical analysis of large-scale datasets to
extract additional patterns and trends~\cite{hastie:mining}. This has
allowed commercial, state, and academic actors to answer questions
which have not previously been possible, due to insufficient data,
analytical techniques, or computational power. Data mining is often
characterised by the use of aggregate data to identify traits and
trends which allow the identification and characterisation of clusters
of people rather than individuals, associations between events, and
forecasting of future events. As such, it has been used in a number of
real-world scenarios such as optimising the layout of retail stores,
attempts to identify disease trends, and mass surveillance.  Many
classical data mining and knowledge discovery applications involve
businesses or marketing~\cite{fayyad:data-mining}, such as clustering
consumers into groups and attempting to predict their behaviour. This
may allow a business to understand their customers and target
promotions appropriately. Such profiling can, however, be used to
characterise individuals for the purpose of denying service when
extending credit, leasing a property, or acquiring insurance. In such
cases, the collection and processing of sensitive data can be
invasive, with significant implications for the individual,
particularly where decisions are made on the basis of inferences that
may not be accurate, and to which the individual is given no right of
reply.  This has become more important of late, as more recent data
mining applications involve the analysis of personal data, much of
which is collected by individuals and contributed to marketers in what
has been termed ``self-surveillance''~\cite{kang:self-surveillance}.
Such personal data have been demonstrated to be highly
valuable~\cite{staiano:money}, and have even been described as the new
``oil'' in terms of the value of their resource~\cite{wef:asset}.
Value aside, such data introduce new challenges for consent as they
can often be combined to create new inferences and profiles where
previously data would have been absent~\cite{heimbach:profile}.

\begin{table}
\caption{Some relevant EU legislation that may apply to data mining
activities}
\label{t:laws}
\begin{tabular}{p{5cm}p{6.5cm}}
\hline\noalign{\smallskip}
Legislation & Some relevant sections \\
\noalign{\smallskip}\svhline\noalign{\smallskip}
Data Protection Directive 95/46/EC & Data processing (Art 1), fair
processing (Art 6(1)), purpose limitation (Art 6(2)), proportionality
(Art 6(3)), consent (Art 7(1)), sensitive data (Art 8) \\
E-Privacy Directive 2002/58/EC as amended by 2009/136/EC & cookies
(Art 5), traffic data (Art 6), location data (Art 9), unsolicited
communication (Art 13) \\
General Data Protection Regulation 2016/679 & Consent (Art 7), right
to be forgotten (Art 17), right to explanation (Art 22), privacy by
design (Art 25) \\
\noalign{\smallskip}\hline\noalign{\smallskip}
\end{tabular}
\end{table}

Data mining activities are legitimised through a combination of legal and
self-regulatory behaviours. In the European Union, the Data Protection
Directive~\cite{eu:directive}, and the forthcoming General Data Protection
Regulation (GDPR) that will succeed it in 2018~\cite{eu:gdpr} govern how data
mining can be conducted legitimately. The e-Privacy Directive also
further regulates some specific aspects of data mining such as cookies
(Table~\ref{t:laws}). In the United States, a self-regulatory
approach is generally preferred, with the Federal Trade Commission offering
guidance regarding privacy protections~\cite{ftc:fip}, consisting of six core
principles, but lacking the coverage or legal backing of the EU's approach.

Under the GDPR, the processing of personal data for any purpose, including
data mining, is subject to explicit opt-in consent from an individual, prior
to which the data controller must explicitly state what data are collected,
the purpose of processing them, and the identity of any other recipients of
the data. Although there are a number of exceptions, consent must generally be
sought for individual processing activities, and cannot be broadly acquired \emph{a priori} for undefined future
uses, and there are particular issues with data mining, transparency
and accountability~\cite{carmichael:discrimination}. Solove~\cite{solove:self-management} acknowledges these
regulatory challenges, arguing that paternalistic approaches are not
appropriate, as these deny people the freedom to consent to
particular beneficial uses of their data. The timing of consent
requests and the focus of these requests need to be managed
carefully; such thinking has also become apparent in the
GDPR.\footnote{e.g., Article 7(3) which allows consent to be withdrawn,
    and Article 17 on the ``right to be forgotten'' which allows
inferences and data to be erased.} The call for
dynamic consent is consistent with Nissenbaum's model of contextual
integrity~\cite{nissenbaum:context},
which posits that all information exchanges are subject to context-specific
norms,
which governs to whom and for what purpose information sharing can be
considered appropriate. When the context is disrupted, perhaps by changing
with whom data are shared, or for what purpose, privacy violations can occur
when this is not consistent with the norms of the existing context. Therefore,
consent can help to uphold contextual integrity by ensuring that if the context
is perturbed, consent is renegotiated, rather than assumed.

Reasoning about how personal data are used has resulted in a new
paradigm, \emph{human-data interaction}, which places humans at the
centre of data flows and provides a framework for studying personal
data collection and use according to three
themes~\cite{mortier:hdi-encyclopedia}:

\begin{itemize}

	\item \emph{Legibility}: Often, data owners are not aware that data mining is even taking place. Even if they are, they may not know what is being collected or analysed, the purpose of the analysis, or the insights derived from it.

	\item \emph{Agency}: The opaque nature of data mining often denies data owners agency. Without any engagement in the practice, people have no ability to provide meaningful consent, if they are asked to give consent at all, nor correct flawed data or review inferences made based on their data.

	\item \emph{Negotiability}: The context in which data are collected and processed can often change, whether through an evolving legislative landscape, data being traded between organisations, or through companies unilaterally changing their privacy policies or practices. Analysis can be based on the linking of datasets derived
	from different stakeholders, allowing insights that no single provider could
	make. This is routinely the case in profiling activities such as credit
    scoring. Even where individuals attempt to obfuscate their data to subvert this practice, it is often possible to re-identify them from such linked data~\cite{narayanan:deanonymizing}. Data owners should have the ability to review how their data are used as circumstances change in order to uphold contextual integrity.

\end{itemize}

Early data protection regulation in the 1980s addressed the increase in
electronic data storage and strengthened protections against unsolicited
direct marketing~\cite{steinke:us-eu}. Mail order companies were able to
develop large databases of customer details to enable direct marketing, or the
trading of such information between companies. When acquiring consent for the
processing of such information became mandatory, such as under the 1984 Data
Protection Act in the UK, this generally took the form of a checkbox on paper
forms, where a potential customer could indicate their willingness for
secondary processing of their data. As technology has evolved away from mail-in
forms being the primary means of acquiring personal information, and the
scope and intent of data protection  moves from regulating direct marketing to
a vast range of data-processing activities, there has been little regulatory
attention paid to how consent is acquired. As such, consent is often acquired
by asking a user to tick a checkbox to opt-in or out of secondary use of their
data. This practice is well-entrenched, where people are routinely asked to
agree to an End-User Licence Agreement (EULA) before accessing software, and
multiple terms of service and privacy policies before accessing online
services, generally consisting of a long legal agreement and an ``I Agree''
button.

A significant body of research concludes that such approaches to acquiring
consent are flawed. Luger et al. find that the terms and conditions provided
by major energy companies are not sufficiently readable, excluding many from
being able to make informed decisions about whether they agree to such
terms~\cite{luger:complexity}. Indeed, Obar and Oeldorf-Hirsch find that the
vast majority of people do not even read such documents~\cite{obar:lie}, with
all participants in a user study accepting terms including handing over their
first-born child to use a social network site. McDonald and Cranor measure the
economic cost of reading lengthy policies~\cite{mcdonald:policies}, noting the
inequity of expecting people to spend an average of ten minutes of their time
reading and comprehending a complex document in order to use a service.
Freidman et~al. caution that simply including more information and more
frequent consent interventions can be counter-productive, by frustrating
people and leading them to making more complacent consent
decisions~\cite{friedman:informed}.

Academic data mining is subject to a different regulatory regime, with fewer
constraints over the secondary use of data from a data protection perspective.
This is balanced by an ethical review regime, rooted in post-war concern over
a lack of ethical rigour in biomedical research. In the US, ethical review for
human subjects research via an institutional review board (IRB) is necessary
to receive federal funding, and the situation is similar in many other
countries. One of the central tenets of ethical human research is to acquire
informed consent before a study begins~\cite{berg:consent}. As such,
institutions have developed largely standardised consent
instruments~\cite{akkad:consent} which researchers can use to meet these
requirements. While in traditional lab-based studies, these consent procedures
can be accompanied by an explanation of the study from a researcher, or the
opportunity for a participant to ask any questions, this affordance is
generally not available in online contexts, effectively regressing to the
flawed EULAs discussed earlier.

 Some of these weaknesses have been examined in the
literature. Hamnes et al. find that consent documents in rheumatological studies
are not sufficiently readable for the majority of the
population~\cite{hamnes:readability}, a finding which is supported by Vu\v
{c}emilo and Borove\v{c}ki who also find that medical consent forms often
exclude important information~\cite{vucemilo:readability}. Donovan-Kicken et al.
examine the sources of confusion when reviewing such
documents~\cite{donovan-kicken:uncertainty}, which include
insufficient discussion of risk and lengthy or overly complex language. Munteanu et~al. examine the ethics approval process in a number of HCI
research case studies, finding that participants often agreed to
consent instruments they have not read or understood, and the rigidity of such
processes can often be at odds with such studies where a ``situational
interpretation'' of an agreed protocol is needed~\cite{munteanu:situational}.
There also lacks agreement among researchers about how to conduct such research
in an ethical manner, with Vitak et~al. finding particular variability regarding
whether data should be collected at large scale without consent, or if acquiring
consent in such cases is even possible~\cite{vitak:belmont}.

Existing means of acquiring consent are inherited from a time when the scope of
data collection and processing was perhaps constrained and could be well
understood. Now, even when the terms of data collection and processing are
understood as written, whether registering for an online service, or
participating in academic research, it is not clear that the form of gaining
the consent was meaningful, or sufficient. Someone may provide consent to
secondary use of their data, without knowing what data this constitutes, who
will be able to acquire it, for what purpose, or when. This is already a
concern when considering the redistribution and processing of self-disclosed
personally identifiable information, but becomes increasingly complex when
extended to historical location data, shopping behaviours, or social network
data, much of which are not directly provided by the individual, and
are nebulous
in scale and content. Moreover concerns may change over time (the
so-called ``privacy paradox''~\cite{barnes:privacy} that has been
demonstrated empirically~\cite{ayalon:retrospective,bauer:temporal}),
which may require changes to previously-granted consent.

Returning to our three themes of \emph{legibility}, \emph{agency}, and
\emph{negotiability}, we can see that:

\begin{itemize}
 	\item Existing EULAs and consent forms may not meet a basic
 	standard of \emph{legibility}, alienating
 	significant areas of the population from understanding what they are being
 	asked to agree to. Furthermore, the specific secondary uses of their data are
 	often not explained.

 	\item EULAs and consent forms are often only used to secure permission once, then
 	often never again, denying people \emph{agency} to revoke their
 	consent when a material change in how their data are used arises.

 	\item Individuals have no power to meaningfully \emph{negotiate} how their
 	data are
 	used, nor to intelligently adopt privacy-preserving behaviours, as they
 	generally do not know which data attributed to them is potentially risky.
\end{itemize}

\section{Case studies}
\label{s:cases}

In this section we examine a number of real-world case studies to
identify instances where insufficient consent mechanisms were
employed, failing to provide people with legibility, agency, and
negotiability.

\subsection{Taste, Ties, and Time}

In 2006, researchers at Harvard University collected a dataset of Facebook
profiles from a cohort of undergraduate students, named ``Tastes, Ties and
Time'' (T3)~\cite{lewis:facebook}. At the time, Facebook considered individual
universities to comprise networks where members of the institution could access the
full content of each other's profiles, despite not having an explicit
friendship with each other on the service. This design was exploited
by having research assistants at the same institution manually
extract the profiles of each member of the cohort.

Subsequently, an anonymised version of the dataset was made publicly
available, with student names and identifiers removed, and terms and
conditions for downloading the dataset made it clear that
deanonymisation was not permitted.  Unfortunately, this proved
insufficient, with aggregate statistics such as the size of the cohort
making it possible to infer the college the dataset was derived from,
and as some demographic attributes were only represented by a single
student, it was likely that individuals could be
identified~\cite{zimmer:public}.

Individuals were not aware that the data collection took place, and did not
consent to its collection, processing, nor subsequent release. As such, this
case falls short in our themes for acceptable data-handling practices:

\begin{itemize}
	\item \textbf{Legibility}: Individuals were not aware their data was
	collected or subsequently released. With a tangible risk of individuals being
	identified without their knowledge, the individual is not in a position to explore any legal
	remedies to hold Facebook or the researchers responsible for any
	resulting harms. In addition, even if consent were sought, it can be difficult
	for individuals to conceptualise exactly which of their data would be included,
	considering the large numbers of
	photos, location traces, status updates, and biographical information a typical
	user might accrue over years, without an accessible means of visualising or
	selectively disclosing these data.

	\item \textbf{Agency}: Without notification, individual users had no way to
	opt-out of the data collection, nor prevent the release of their data. As a
	side-effect of Facebook's university-only network structure at the time, the
	only way for
	somebody to avoid their data being used in such a manner was to leave these institution networks, losing much of the utility of the service in the process. This parallels
	Facebook's approach to releasing
	other products, such as the introduction of News Feed in 2006. By
	broadcasting profile updates to one's network, the effective visibility of
	their data was substantially increased, with no way to opt-out of the feature
	without leaving the service entirely. This illusory loss of control was
	widely criticised at the time~\cite{hoadley:newsfeed}.

	\item \textbf{Negotiability}: In this respect, the user's relationship with
	Facebook itself is significant. In addition to IRB approval, the study was conducted with
	Facebook's
	permission~\cite{lewis:facebook}, but Facebook's privacy policy at the
	time did not allow for Facebook to
	share their data with the researchers.\footnote{Facebook Privacy Policy,
	February 2006: \url
	{http://web.archive.org/web/20060406105119/http://www.facebook.com/policy.php}}
	Therefore, the existing context for sharing information on Facebook was
	disrupted
	by this study. This includes the normative expectation that data are shared
	with
	Facebook for the purpose of sharing information with one's social network, and
	not myriad third parties. In addition, no controls were extended to the
	people involved to prevent it from happening, or to make a positive decision to
	permit this new information-sharing context.

\end{itemize}

\subsection{Facebook emotional contagion experiment}
In 2012, researchers at Facebook and Cornell University conducted a
large-scale
experiment on 689,003 Facebook users. The study manipulated the presentation
of stories in Facebook's News Feed product, which aggregates recent content
published by a user's social network, to determine whether biasing the
emotional content of the news feed affected the emotions that people expressed in
their own disclosures~\cite{kramer:contagion}.

While the T3 study highlighted privacy risks of nonconsensual data sharing, the
emotional contagion experiment raises different personal risks from
inappropriate data mining activities. For example, for a person suffering from
depression, being subjected to a news feed of predominantly
depressive-indicative content could have catastrophic consequences, particularly
considering the hypothesis of the experiment that depressive
behaviour would increase under these circumstances. Considering the
scale at which the
experiment
was conducted, there was no mechanism for excluding such vulnerable people,
nor measuring the impact on individuals to mitigate such harms. Furthermore,
as the study was not age-restricted, children may have unwittingly been
subjected to the study~\cite{hill:facebook}. Rucuber notes that the harms to
any one individual in such experiments can be masked by the scale of the
experiment~\cite{recuber:milgram}.

Beyond the research context, this case highlights the broader implications of
the visibility of media, whether socially-derived or from mainstream media,
being algorithmically controlled. Napoli argues that this experiment highlights
Facebook's ability to shape public discourse by altering the news feed's
algorithm to introduce political bias~\cite{napoli:governance}, without any
governance to ensure that such new media are acting in the public interest. The
majority of Facebook users do not know that such filtering happens at all, and
the selective presentation of content from one's social network can cause
social repercussions where the perception is that individuals are withholding
posts from someone, rather than an algorithmic intervention by
Facebook~\cite{eslami:algorithms}.

This case shows one of the greater risks of opaque data mining. Where people are
unaware such activities are taking place, they lose all power to act
autonomously to minimise the risk to themselves, even putting aside the
responsibility of the
researchers in this instance. We now consider how this case meets our three core
themes:

\begin{itemize}

\item \textbf{Legibility}: Individuals were unaware that they were participants
in the research. They would have no knowledge or
understanding of the algorithms which choose
which content is presented on the news feed, and how they were altered for this
experiment, nor that the news feed is anything other than a chronological
collection of content provided by their social network. Without this insight,
the cause of a perceptible change in the emotional bias in the news feed can not
be reasoned. Even if one is aware that the news feed is algorithmically
controlled,
without knowing which data are collected or used in order to determine
the relevance of individual stories, it is difficult to reason why certain
stories are displayed.

\item \textbf{Agency}: As in the T3 case, without awareness of the experiment
being conducted, individuals were unable to provide consent, nor opt-out of the
study. Without an understanding of the algorithms which drive the news feed, nor
how they were adjusted for the purposes of this experiment, individuals are
unable to take actions, such as choosing which information to disclose or hide
from Facebook in an effort to control the inferences Facebook makes, nor to
correct any inaccurate inferences. At the most innocuous level, this might be
where Facebook has falsely inferred a hobby or interest, and shows more content
relating to that. Of greater concern is when Facebook, or the researchers
in this study, are unable to detect when showing more
depressive-indicative content could present a risk.

\item \textbf{Negotiability}: In conducting this study, Facebook unilaterally
changed the relationship its users have with the service, exploiting those who
are unable to control how their information is used~\cite{selinger:co-opted}.
At the time of the study, the Terms of Service to which users agree when
joining Facebook did not indicate that data could be used for research
purposes~\cite{hill:facebook}, a clause which was added after the data were
collected. As a commercial operator collaborating with academic researchers,
the nature of the study was ambiguous, with Facebook having an internal
product improvement motivation, and  Cornell researchers aiming to contribute
to generalisable knowledge. Cornell's IRB deemed that they did not need to
review the study because Facebook provided the data,~\footnote{Cornell statement: \url{https://perma.cc/JQ2L-TEXQ}},
but
the ethical impact on the unwitting participants is not dependent on who collected the data. As Facebook has no legal requirement to
conduct an ethical review of their own research, and without oversight from
the academic collaborators, these issues did not surface earlier. Facebook has
since adopted an internal ethics review process~\cite{jackman:irb}, however it
makes little reference to mitigating the impact on participants, and mostly
aims to maximise benefit to Facebook. Ultimately, these actions by researchers
and institutions with which individuals have no prior relationship serve to
disrupt the existing contextual norms concerning people's relationship with
Facebook, without extending any ability to renegotiate this relationship.

\end{itemize}

\subsection{NHS sharing data with Google}
In February 2016, the Google subsidiary DeepMind
announced a collaboration with the National Health Service's Royal Free London
Trust to build a mobile application titled Streams to support the detection of
acute kidney
injury (AKI) using machine learning techniques. The information sharing
agreement permitting this collaboration gives DeepMind ongoing and historical
access to all identifiable patient data collected by the trust's three
hospitals~\cite{hodson:google}.

While the project is targeted at supporting those at risk of AKI, data
relating to all patients are shared with DeepMind, whether they are at
risk or not. There is no attempt to gain
the consent of those patients, or to provide an obvious opt-out mechanism. The
trust's privacy policy only allows data to be shared without explicit consent
for ``your direct care''.~\footnote{Royal Free London Trust Privacy Statement:
    \url{https://perma.cc/33YE-LYPF}}
Considering that Streams is
only
relevant to
those being
tested for kidney disease, it follows that for most people, their data are
collected and processed without any direct care benefit~\cite{hodson:approval},
in violation of this policy.  Given the diagnostic purpose of the app,
such an
application could constitute a medical device, however no regulatory approval
was sought by DeepMind or the trust~\cite{hodson:approval}.

Permitting private companies to
conduct data mining within the medical domain disrupts existing norms, by
occupying a space that lies between direct patient care and academic research.
Existing ethical approval and data-sharing regulatory mechanisms have not been
employed, or are unsuitable
for properly evaluating the potential impacts of such work. By not limiting
the scope of the data collection nor acquiring informed
consent, there is no opportunity for individuals to protect
their data. In addition, without greater awareness of the collaboration, broader
public debate about
the acceptability of the practice is avoided, which is of importance
considering the sensitivity of the data involved. Furthermore, this fairly
limited
collaboration can normalise a broader sharing of data in the future, an
eventuality which is more likely given an ongoing strategic partnership forged
between DeepMind and the trust~\cite{hodson:approval}.
We now consider this case from the perspective of our three themes:

\begin{itemize}

\item \textbf{Legibility}: Neither patients who could directly benefit from
improved detection of AKI, nor all other hospital patients, were aware that
their data were being shared with DeepMind.
Indeed, this
practice in many data mining activities -- identifying patterns to produce
insight from myriad data -- risks violating a fundamental principle of
data protection regulation; that of proportionality~\cite{eu:gdpr}.

\item \textbf{Agency}: The NHS collects data from its functions in a number of
databases, such as the
Secondary Uses Service which provides a historical record of treatments in the
UK, and can be made available to researchers. Awareness of this
database and its research purpose is mostly constrained to leaflets and posters
situated in GP practices. If patients wish to opt-out of
their data being used they must insist on it, and are likely to be reminded of
the public health benefit and discouraged from opting-out~\cite{brown:consent}.
Without being able to assume awareness of the SUS, nor individual consent being
acquired, it is difficult for individuals to act with agency. Even assuming
knowledge of this collaboration, it would require particular understanding of
the functions of the NHS to know that opting-out of the SUS would limit
historical treatment data made available to DeepMind. Even where someone is
willing to share their data to support their direct care, they may wish to
redact information relating to particularly sensitive diagnoses or treatments,
but have no mechanism to do so.

\item \textbf{Negotiability}: The relationship between patients and their
clinicians is embodied in complex normative expectations of confidentiality
which are highly context-dependent~\cite{sankar:confidentiality}. Public
understanding of individual studies is already low~\cite{brown:consent}, and the
introduction of sophisticated data mining techniques into the diagnostic
process which existing regulatory mechanisms are not prepared for disrupts
existing norms around confidentiality and data sharing. The principle of
negotiability holds that patients should be able to review their willingness
to share data as their context changes, or the context in which the data are
used. Existing institutions are unable to uphold this, and the solution may
lie in increased public awareness and debate, and review of policy and regulatory
oversight to reason a more appropriate set of norms.

\end{itemize}

How each of these case studies meets the principles of legibility, agency, and
negotiability is summarised in Table~\ref{t:hdi}.

\begin{table}[htp]
\bgroup
\def\arraystretch{1.5}%
\setlength{\tabcolsep}{0.5em}
\begin{tabular}{| p{0.3\linewidth} | p{0.15\linewidth} |
      p{0.5\linewidth} |}

\hline \textbf{Case study} & \mbox{\textbf{HDI principle}} & \textbf{How was principle violated?}\\ \hline

\multirow{3}{*}{Taste, Ties and Time} & Legibility & Individuals unaware of data collection\\\cline{2-3}
& Agency & No way to opt-out of data collection\\\cline{2-3}
& Negotiability & Data collection violated normative expectation with Facebook\\ \hline

\multirow{3}{0.3\textwidth}{Facebook emotional contagion experiment} & Legibility & Individuals unaware they were participants\\\cline{2-3}
& Agency & No way to opt-out of participation\\ \cline{2-3}
& Negotiability & Research not permitted by Facebook's terms\\ \hline

\multirow{3}{0.3\textwidth}{NHS sharing data with Google} & Legibility & Patients unaware of data sharing\\ \cline{2-3}
& Agency & Poor awareness of secondary uses of data and difficulty of opting out\\ \cline{2-3}
& Negotiability & Data mining can violate normative expectations of medical confidentiality\\ \hline

\end{tabular}
\egroup

\caption{Summary of how the three case studies we examine violate the principles of human-data interaction}
\label{t:hdi}
\end{table}

\section{Alternative consent models}
\label{s:consent}

In Section~\ref{s:bg} we discussed some of the shortcomings with existing means
of acquiring consent for academic research and commercial services including
data mining, and discussed three case studies in
Section~\ref{s:cases}. Many of the concerns in these case studies
revolved around an inability to provide or enable consent on the part
of participants. We now discuss the state-of-the-art in providing
meaningful consent for today's data-mining activities.

The acquisition of informed consent can broadly be considered
to be \emph{secured} or \emph{sustained} in nature~\cite{luger:informed}.
Secured consent encompasses the forms we discussed in
Section~\ref{s:bg}, where consent is gated by a single EULA or consent
form at the beginning of the data collection process and not
revisited. Conversely, sustained consent involves ongoing
reacquisition of consent over the period that the data are collected
or used.  This might mean revisiting consent when the purpose of the
data collection or processing has changed, such as if data are to be
shared with different third parties, or if the data subject's context
has changed. Each interaction can be viewed as an individual consent
\emph{transaction}~\cite{miller:consent-transactions}. In research, this can also mean extending more granular
control to participants over which of their data are collected, such
as in Sleeper et al.'s study into self-censorship behaviours on
Facebook, where participants could choose which status updates they were
willing to share with researchers~\cite{sleeper:censor}. This approach has a number of
advantages. Gaining consent after the individual has had experience
with a particular service or research study may allow subjects to make
better-informed decisions than a sweeping form of secured consent.
Furthermore, sustained consent can allow participants to make more
granular decisions about what they would be willing to share, with a
better understanding of the context, rather than a single consent form
or EULA being considered \emph{carte blanche} for unconstrained data
collection.

The distinction between secured and sustained consent reveals a tension between
two variables: \emph{burden} -- the time spent and cognitive load required to
negotiate the consent process, and \emph{accuracy} -- the extent to which the
effect of a consent decision corresponds with a person's expectations. While a
secured instrument such as a consent form minimises the burden on the
individual, with only a single form to read and comprehend, the accuracy is
impossible to discern, with no process for validating that consent decision in
context, nor to assess the individual's comprehension of what they have agreed
to. Conversely, while a sustained approach -- such as asking someone whether
they are willing for each item of personal data to be used for data mining
activities -- may improve accuracy, the added burden is significant and can be
frustrating, contributing to attrition, which is particularly problematic in
longitudinal studies~\cite{hektner:esm}.

In some domains other than data mining, this distinction has already been
applied. In biomedical research, the consent to the use of samples is commonly
distinguished as being broad or dynamic. Broad consent allows samples to be used
for a range of experiments within an agreed framework without consent being
explicitly required~\cite{steinsbekk:dynamic}, whereas dynamic consent involves
ongoing engagement with participants, allowing them to see how their samples are
used, and permitting renegotiation of consent if the samples are to be used for
different studies, or if the participant's wishes change~\cite{kaye:dynamic}.
Despite the differences from the data mining domain, the same consent
challenges resonate.

Various researchers have proposed ways of minimising the burden of
consent, while simultaneously collecting meaningful and accurate
information from people. Williams et~al. look at sharing medical data,
enhancing agency with a dynamic
consent model that enables control of data electronically, and improved
legibility by providing patients with information about how their data are
used~\cite{williams:consent}. Gomer et~al. propose the use of agents who make
consent decisions
on behalf of individuals to reduce the burden placed on them, based on
preferences they have expressed which are periodically
reviewed~\cite{gomer:agents}. Moran et~al. suggest that consent can be
negotiated in multi-agent environments by identifying interaction patterns to
determine appropriate times to acquire consent~\cite{moran:agent}.

We have discussed legibility as an important aspect of HDI, and
Morrison et~al. study how to visualise collected data to research
participants~\cite{morrison:personalised-representations}.
Personalised visualisations led participants in an empirical study to
exit the study earlier, which might mean that secured consent was leading
participants to continue beyond the appropriate level of data
collection. In much earlier work, Patrick looked at presenting user
agreements contextually (rather than at the beginning of a
transaction as in secured consent) and developed a web-based widget to
do so~\cite{patrick:agreements}.

Such dynamic approaches to consent are not universally supported. Steinsbekk
et~al. suggest that where data are re-used for multiple studies, there is no
need to acquire consent for each one where there are no significant ethical
departures because of the extra burden, arguing that it puts greater
responsibility on individuals to discern whether a study is ethically
appropriate than existing governance structures~\cite{steinsbekk:dynamic}.

In previous work, we have developed a method for acquiring consent which aims to
maximise accuracy and minimise burden, satisfying both requirements, bringing
some of the principles of dynamic consent to the data mining
domain~\cite{hutton:consent}, while aiming to maintain contextual integrity by
respecting prevailing social norms.
While many of the consent approaches discussed in this chapter may satisfy a
legal requirement, it is not clear that this satisfies the expectations of
individuals or society at large, and thus may violate contextual integrity.

In a user study, we examine whether prevailing norms representing willingness to
share specific types of Facebook data with researchers, along with limited data
about an individual's consent preferences, can be used to minimise burden
and maximise accuracy. The performance of these measures were compared to two
controls: one for accuracy and for burden. In the first instance,
a sustained consent process which gains permission for the use of each
individual item of Facebook data would maximise accuracy while pushing great
burden on to the individual. Secondly, a single consent checkbox minimises
burden, while also potentially minimising the accuracy of the method. The
contextual integrity method works similarly to this approach, by asking a series
of consent questions until the individual's conformity to the social norm can be
inferred, at which point no more questions are asked.

\begin{figure}[tbp]
    \centerline{\resizebox{0.8\linewidth}{!}{\includegraphics{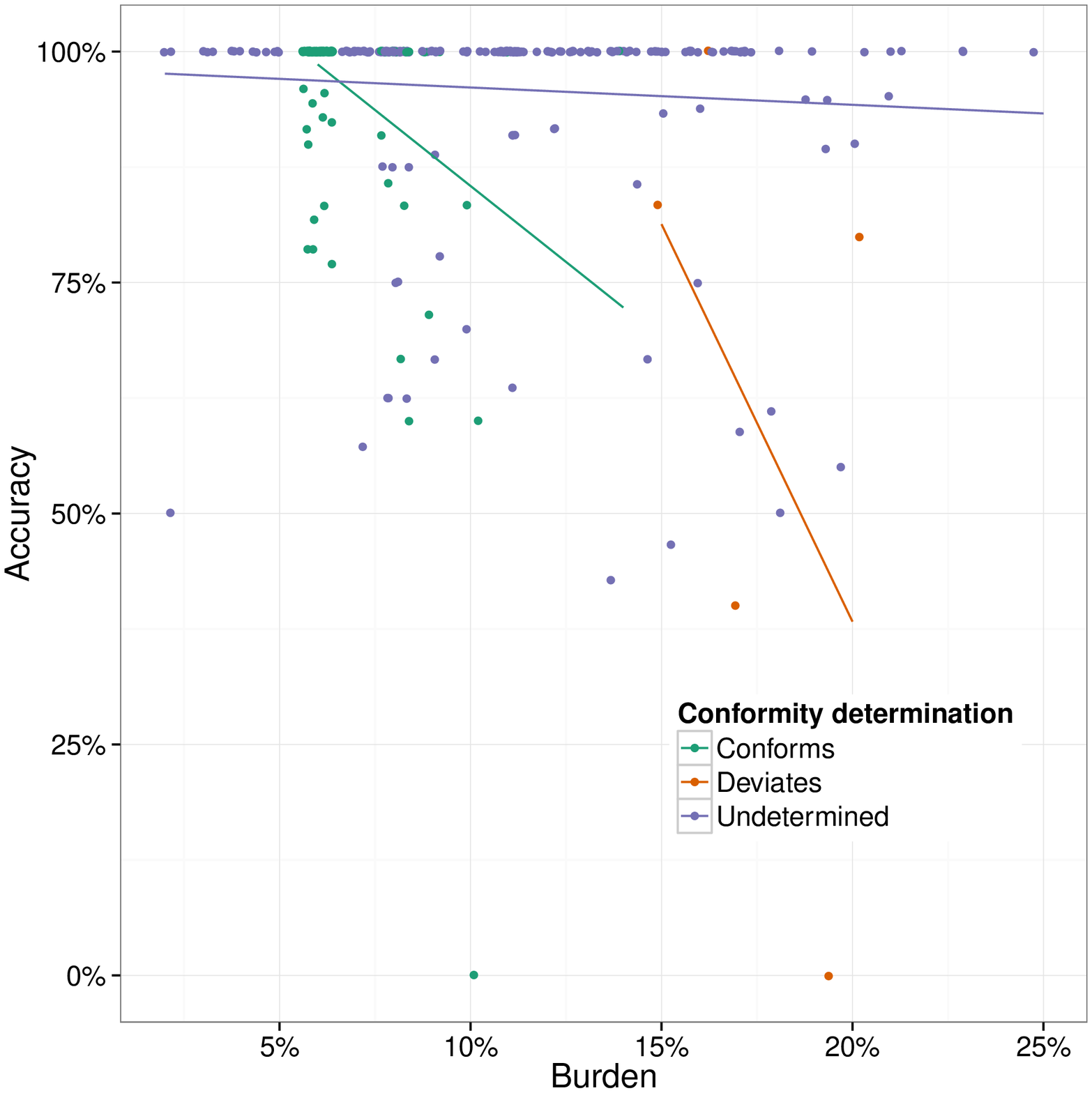}}}
    \caption{\label{p:consent-normacc}
Scatterplot showing the relationship between norm conformity and accuracy. As
indicated by the cluster after 5 questions (5\% burden), high accuracy can be
preserved when
norm conformity is detected quickly, although the technique is not useful for
people who are highly norm deviant. Note that the points have been jittered to
improve readability.~\cite{hutton:consent}}
\end{figure}

For 27.7\% of participants, this method is able to achieve high accuracy of
96.5\% while reducing their burden by an average of 41.1\%. This is highlighted
in Figure~\ref{p:consent-normacc}, showing a cluster of norm-conformant
participants achieving high accuracy and low burden. This indicates that
for this segment of the population, the contextual integrity approach both
improves accuracy and reduces burden compared to their respective controls.
While this does indicate the approach is not suitable for all people, norm
conformity is able to be quickly determined within six questions. Where one
does not conform to the norm, the sustained approach can be automatically used
as a fallback, which maintains accuracy at the cost of a greater time burden
on the individual. Even in less optimal cases, the technique can reduce
the burden by an average of 21.9\%.

While the technique assessed in this user study is prototypical in its nature,
it highlights the potential value of examining alternative means of acquiring
consent, which has seen little innovation in both academic and commercial
domains. Moreover, while this technique is not universally applicable, this only
highlights that the diversity of perspectives, willingness to engage, and
ability to comprehend consent language requires a plurality of approaches.

\section{Discussion}
In this chapter we have illustrated how data mining activities, in both
academic and commercial contexts, are often opaque by design. Insufficient
consent mechanisms can prevent people from understanding what they
are agreeing to, particularly where the scope of the data collected or with
whom it is shared is changed without consent being renegotiated. Indeed, as
in our three case studies, consent is often not sought at all. 

We have considered the impacts of opaque data mining in terms of legibility,
agency, and negotiability. We now propose some best practices for
conducting data mining which aim to satisfy these three themes.

\subsection{Legibility}
In order to make data mining more acceptable, it is not sufficient to simply
make processes more transparent. Revealing the algorithms, signals, and
inferences
may satisfy a particularly technically competent audience, but for most people
does not help them understand what happens to their data, in order to make an
informed decision over whether to consent, or how they can act with any agency.

The incoming General Data Protection Regulation (GDPR) in the European Union
requires consent language to be concise, transparent, intelligible and easily
accessible~\cite{eu:gdpr}, which as indicated in the literature, is currently
not a universal practice. As highlighted in our three case studies, the
absence of any meaningful consent enabling data to be used beyond its original
context, such as a hospital or social network site, is unacceptable. Even
without adopting more sophisticated approaches to consent as discussed in
Section~\ref{s:consent}, techniques to notify and reacquire consent such that
people are aware and engaged with ongoing data mining practices can be
deployed. As discussed earlier, a practical first step is to ensure all
consent documents can be understood by a broad spectrum of the population.

\subsection{Agency}
Assuming that legibility has been satisfied, and people are able to understand
how their data are being used, the next challenge is to ensure people are
able to act autonomously and control how their data are used beyond a single
consent decision. Some ways of enabling this include ensuring people can
subsequently revoke their consent for their data to be used at any time, without
necessarily being precipitated by a change in how the data are used. In the
GDPR, this is enshrined through the right to be
forgotten~\cite{eu:gdpr} that 
includes the cascading revocation of data between data controllers.

Legibility can also enable agency by allowing people to act in a certain way in
order to selectively allow particular inferences to be made. By being able to
choose what they are willing to share with a data collector in order to satisfy
their own utility, some of the power balance can be restored, which has been
previously tipped
towards the data collector who is able to conduct analyses at a scale beyond any
individual subject's capabilities.

\subsection{Negotiability}
As discussed in Section~\ref{s:consent}, Nissenbaum's contextual
integrity~\cite{nissenbaum:context} can be used to detect privacy violations
when the terms of data-handling have changed in such a way that existing norms
are breached. The principle of negotiability is key to preventing this, by
allowing people to make ongoing decisions about how their data are used as
contexts evolve, whether their own, environmentally, or that of the data
collector.

Dynamic consent in the biobanks context~\cite{kaye:dynamic}
could be adapted to
allow data subjects to be notified and review how their data are being used,
whether for new purposes or shared with new actors, allowing consent to be
renegotiated. Our consent method informed by contextual
integrity~\cite{hutton:consent} is one such approach which aims to tackle this
problem, by allowing people to make granular consent decisions without being
overwhelmed. Adopting the principles of the GDPR, which emphasises dynamic
consent, can support negotiability, with
guidance made available for organisations wishing to apply these
principles~\cite{tankard:gdpr}.

\section{Conclusion}

Data mining is an increasingly pervasive part of daily life, with the
large-scale collection, processing, and distribution of personal data being
used for myriad purposes. In this chapter, we have outlined how this often
happens without consent, or the consent instruments used are overly complex or
inappropriate. Data mining is outgrowing existing regulatory and ethical
governance structures, and risks violating entrenched norms about the
acceptable use of personal data, as illustrated in case studies spanning the
commercial and academic spheres. We argue that organisations involved in data
mining should provide legible consent information such that people can
understand what they are agreeing to, support people's agency by allowing them
to selectively consent to different processing activities, and to support
negotiability by allowing people to review or revoke their consent as the
context of the data mining changes. We have discussed recent work which
dynamically negotiates consent, including a technique which leverages social
norms to acquire granular consent without overburdening people. We call for
greater public debate to negotiate these new social norms collectively, rather
than allowing organisations to unilaterally impose new practices without
oversight.

\section{Acknowledgements}

This work was supported by the Engineering and Physical Sciences Research Council [grant number EP/L021285/1].

\bibliographystyle{spmpsci}

\end{document}